\documentclass[letterpaper, 10 pt, conference]{ieeeconf}  % 
\IEEEoverridecommandlockouts                              % d
\title{\LARGE \bf Computation of Feasible Assume-Guarantee Contracts:\\ A Resilience-based Approach}
\author{Negar Monir$^{1}$, Youssef AIT SI$^{2}$, Ratnangshu Das$^{3}$,\\
Pushpak Jagtap$^{3}$, Adnane Saoud$^{2}$, and Sadegh Soudjani$^{4}$ % <-this % stops a space
\thanks{The research of N. Monir is supported by the EPSRC EP/W524700/1 and Newcastle University Global Scholarship. The research of S. Soudjani is supported by the following grants: EIC 101070802 and ERC 101089047. The research of R. Das and P. Jagtap is supported by ARTPARK and Siemens.}
\thanks{$^{1}$ N. Monir is with the School of Computing, Newcastle University, United Kingdom
        {\tt\small (s.seyedmonir2@ncl.ac.uk)}}
\thanks{$^{2}$ Y. Ait Si and A. Saoud are with the University Mohammed IV Polytechnic (UM6P), Morocco {\tt\small \{youssef.aitsi, adnane.saoud\}@um6p.ma}}%
\thanks{$^{3}$ R. Das and P. Jagtap are with the Centre for Cyber-Physical Systems, IISc, Bangalore, India {\tt\small \{ratnangshud,pushpak\}@iisc.ac.in}}%
\thanks{$^{4}$ S. Soudjani is with the Max Planck Institute for Software Systems, Germany, and the University of Birmingham, United Kingdom (\tt\small sadegh@mpi-sws.org)}%
}
\usepackage[usenames,dvipsnames,svgnames,table]{xcolor}
\definecolor{mediumblue}{rgb}{0.0, 0.0, 0.8}
\definecolor{mediumcandyapplered}{rgb}{0.89, 0.02, 0.17}
\definecolor{nazar}{rgb}{0.7, 0.5, 0.9}
\makeatletter
\let\NAT@parse\undefined
\makeatother
\usepackage[colorlinks=true, citecolor=mediumblue, linkcolor=mediumblue, urlcolor = nazar, final]{hyperref}
\usepackage{cite}
\usepackage{amsmath,amssymb,amsfonts}
\usepackage{graphicx}
\usepackage{textcomp}
\usepackage{color}
\usepackage{tikz}
\usepackage{xspace}
\usepackage[many]{tcolorbox}
\usepackage[linesnumbered,ruled]{algorithm2e}

\usepackage{algorithm}
\usepackage{algpseudocode}
\usepackage{wrapfig}
\newcommand{\ball}{\Omega}
\def\reals{\mathbb{R}}

\newtheorem{theorem}{Theorem}[section]

\newtheorem{definition}{Definition}[section]
\newtheorem{remark}{Remark}
\newtheorem{problem}[theorem]{Problem}
\newtheorem{example}{Example}

\newcommand{\always}{\Box}
\newcommand{\eventually}{\Diamond}

\newcommand{\nex}{\mathord{\bigcirc}}

\newtcolorbox{resp1}[1][]{%
	enhanced jigsaw,%
	colback=gray!5!white,%
	colframe=gray!80!black,%
	size=small,%
	boxrule=1pt,%
	halign title=flush center,%
	coltitle=black,%
	breakable,%
	drop shadow=black!50!white,%
	attach boxed title to top left={xshift=1cm,yshift=-\tcboxedtitleheight/2,yshifttext=-\tcboxedtitleheight/2},%
	minipage boxed title=3cm,%
	boxed title style={%
		colback=white,%
		size=fbox,%
		boxrule=1pt,%
		boxsep=2pt,%
		underlay={%
			\coordinate (dotA) at ($(interior.west) + (-0.5pt,0)$);
			\coordinate (dotB) at ($(interior.east) + (0.5pt,0)$);
			\begin{scope}[gray!80!black]
				\fill (dotA) circle (2pt);
				\fill (dotB) circle (2pt);
			\end{scope}
		}%
	},%
	#1%
}

\makeatletter
\newcommand*{\rom}[1]{\expandafter\@slowromancap\romannumeral #1@}
\makeatother

\begin{document}
\maketitle
\thispagestyle{empty}
\pagestyle{empty}
%%%%%%%%%%%%%%%%%%%%%%%%%%%%%%%%%%%%%%%%%%%%%%%%%%%%%%%%%%%%%%%%%%%%%%%%%%%%%%%%
\begin{abstract}
We propose a resilience-based framework for computing feasible assume-guarantee contracts that ensure the satisfaction of temporal specifications in interconnected discrete-time systems. Interconnection effects are modeled as structured disturbances. We use a resilience metric, the maximum disturbance under which local specifications hold, to refine assumptions and guarantees across subsystems iteratively. We first demonstrate correctness and monotone refinement of guarantees for two subsystems. Then, we extend our approach to general networks of \( L \) subsystems using weighted combinations of interconnection effects. We instantiate the framework on linear systems by meeting finite-horizon safety, exact-time reachability, and finite-horizon reachability specifications, and on nonlinear systems by fulfilling general finite-horizon specifications.  Our approach is demonstrated through numerical linear examples and a nonlinear DC microgrid case study, showcasing the impact of our framework on verifying temporal logic specifications with compositional reasoning.
\end{abstract}
%%%%%%%%%%%%%%%%%%%%%%%%%%%%%%%%%%%%%%%%%%%%%%%%%%%%%%%%%%%%%%%%%%%%%%%%%%%%%%%%
\section{INTRODUCTION}
\noindent\textbf{Motivation.} 
Interconnected cyber–physical systems, such as power networks, transportation systems, and industrial automation, must satisfy temporal specifications despite the mutual influences among subsystems.
Assume-Guarantee (AG) reasoning offers a powerful compositional framework to address these challenges by allowing decentralized verification based on local specifications and interconnection assumptions \cite{adnane2018AGC, saoud2019compositional, sharf2021assume, garatti2024, alaoui2024contract}. However, a significant difficulty lies in identifying feasible assumptions that ensure specifications are met throughout the network. To address this issue, we adopt the resilience-based perspective proposed in \cite{saoud2023temporal, saoud2024temporal}: by measuring the maximum disturbance a subsystem can tolerate while still meeting its specifications, we establish a principled way to refine assumptions and guarantees across interconnected subsystems. 
\smallskip

\noindent\textbf{Related Works.} Assume-Guarantee Contracts (AGCs) have a long history as a framework for modular reasoning in system design, originally developed in the software verification community. The initial foundations were established by Abadi and Lamport in the 1990s \cite{lamport1993composing}, in which AG-style reasoning has been applied to reactive modules. The framework has been later formalized by \cite{benveniste2007multiple} for system-level design, which involved contracts consisting of assumptions about the environment and guarantees regarding component behavior. AG reasoning is extended by \cite{benveniste2018contracts} to define formal operations on contracts, such as composition, satisfaction, and refinement, facilitating hierarchical reasoning in heterogeneous systems.

In the realm of continuous-time systems, the work by \cite{adnane2018AGC} has introduced two semantics—weak and strong satisfaction semantics—for AGCs. The authors have demonstrated compositionality results regarding invariance properties in both directed acyclic and cyclic interconnections. This research has been further expanded by \cite{saoud2021assume}, in which general properties beyond invariance have been explored, and the relationship between satisfaction strength and network structure has been clarified. Similar results have been presented in \cite[Section $2$]{saoud2019compositional} for discrete-time systems. Additionally, \cite{benveniste2018contracts} has approached the issue from the perspective of meta-theory by proposing contracts for verifying system properties, specifically for continuous-time systems. Later, \cite{girard2022invariant} has established a link between AGC satisfaction and the presence of invariant sets over finite horizons, enabling verification through set-theoretic techniques. Concurrently, \cite{chen2019compositional} has investigated the compositionality results for networked systems based on set invariance, developing a scalable verification framework that has utilized AG reasoning and set-theoretic computations to characterize invariant sets in interconnected systems. More recently, satisfaction of signal temporal logic tasks has been studied in \cite{liu2022compositional,kordabad2025data} using AG reasoning based on funnel-based methods and barrier certificates.

A significant advancement in computational tools for AGCs has been made by \cite{sharf2021assume}, in which a discrete-time framework has been provided with formal notions of satisfaction, refinement, and cascaded composition. For contract synthesis, \cite{ghasemi2020compositional} introduces a convex parameterization for linear systems with time-varying safety. Complementary approaches include Pareto-based \cite{zonetti2019symbolic}, epigraph-based \cite{chen2019compositional}, and quantitative \cite{eqtami2019quantitative} methods. Building on these ideas, distributed negotiation has been proposed to iteratively refine local contracts toward global safety in \cite{tan2024contract}. Notably, these methods have concentrated mainly on safety specifications.

A complementary area of research has focused on defining resilience metrics to quantify how well dynamical systems can withstand disturbances while still meeting temporal logic specifications. Classical robustness approaches, such as those developed by \cite{fainekos2009robustness, donze2010robust}, assess how far a trajectory remains from violating a specification, often using space-time distance measures. These concepts have since been extended through methods like discounting \cite{almagor2014discounting}, averaging semantics \cite{bouyer2014averaging}, and probabilistic interpretations for stochastic systems \cite{ilyes2023stochastic, farahani2018shrinking}. However, traditional measures of robustness typically evaluate signal traces without explicitly considering system dynamics. In contrast, \cite{saoud2023temporal} has introduced a formal resilience metric defined as the maximum disturbance magnitude that allows the system to satisfy a temporal logic formula. This concept is further extended to compute resilience for linear temporal logic specifications \cite{saoud2024temporal}, synthesize maximally resilient controllers \cite{ait2025maximal}, and assess resilience of power systems \cite{monir2025logic} and water resource recovery facilities \cite{laino2025logic}.
These works establish resilience as a powerful tool for disturbance-aware verification; however, they have not been employed for AGC refinement, which is the focus of this paper.
\smallskip

\begin{figure}
    \centering
    \includegraphics[width=.9\linewidth]{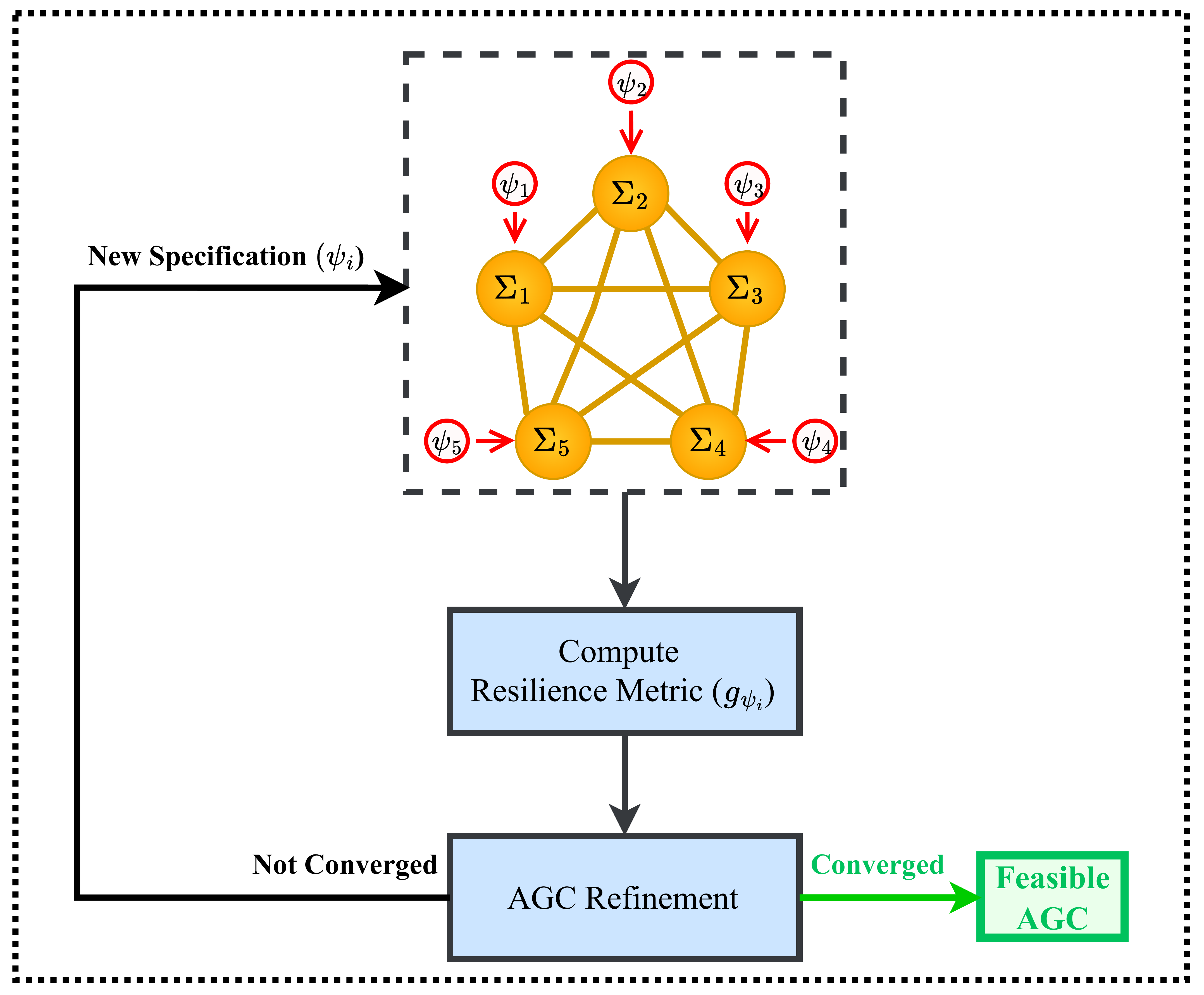}
    \caption{High-level representation of the proposed framework. Each subsystem $i$ receives internal inputs from neighbors and must meet a local specification~$\psi_i$. The process starts by computing a resilience metric that indicates the maximum internal input each subsystem can handle while satisfying $\psi_i$. These metrics are used to iteratively refine AGCs across subsystems until convergence.}
    \label{fig:high level}
\end{figure}

\noindent\textbf{Contributions.} We utilize the resilience metric proposed by \cite{saoud2024temporal} to compute feasible AGCs for verifying temporal specifications in interconnected discrete-time systems, treating internal inputs from neighboring subsystems as structured disturbances. Building on this, we introduce a resilience-theoretic perspective to AG reasoning. This leads to a scalable algorithm for contract refinement. We first define AGCs based on the maximum disturbance each subsystem can tolerate while meeting its local temporal specification. For two subsystems, we propose an iterative algorithm that alternates between computing assumptions and guarantees based on resilience, and prove its correctness and the monotonicity of assumption updates. Next, we extend this approach to general networks of $L$ subsystems by aggregating interconnection effects through weighted disturbance models, facilitating compositional reasoning in large-scale scenarios. The proposed approach is demonstrated for various specifications on both linear and nonlinear dynamics. A high-level representation of our approach is shown in Fig.~\ref{fig:high level}. 
The proofs of statements are relegated to the extended version \cite{monir2025AGC}. 
% The proofs of statements are relegated to the extended version \cite{monir2025AGC}.
The proofs of statements are relegated to the appendix.

%%%%%%%%%%%%%%%%%%%%%%%%%%%%%%%%%%%%%%%%%%%%%%%%%%%%%%%%%%%%%%%%%%%%%%%%%%%%%%%%
\section{Preliminaries and Problem Formulation} \label{sec pre}
%%%%%%------------------- Notations ------------------%%%%
\noindent\textbf{Notations.} The symbols $\reals$, $\reals_{\geq 0}$, $\mathbb{N}$, and $\mathbb{N}_{\geq n}$ denote the sets of real, nonnegative real numbers, nonnegative
integers, and integers greater than or equal to $n\in\mathbb{N}$, respectively.
We also use the notation $\mathbb{N}_{n} := \{0,1,2,\ldots,n\}$ for any $n\in \mathbb{N}$.
We use $\reals^{n\times m}$ to denote the space of real matrices
with $n$ rows and $m$ columns.
For a matrix $A\in \reals^{n\times m}$, $A^\top$ represents the transpose of $A$. For a vector $x\in\reals^n$, we use $\|x\|$ and $\|x\|_\infty$ to denote the Euclidean and infinity norm, respectively. 
We use $\mathbb{I}$ to denote the identity matrix. 
An interval in $\mathbb{R}^n$ is a set denoted by $X=[\underline{x}_1,\overline{x}_1]\times \ldots \times [\underline{x}_n,\overline{x}_n]$ and defined as $X=\left\{x \in \mathbb{R}^n \mid \underline{x}_i \leq x_i \leq \overline{x}_i,~i\in \{1,2,\ldots,n\}\right\}$, where $x_i \in \mathbb{R}$ represents the $i^{th}$ component of the vector $x \in \mathbb{R}^n$. In particular, when $\underline{x}_i=\underline{x}$ and $\overline{x}_i=\overline{x}$ for all $i \in \{1,2,\ldots,n\}$, then the interval $X$ can be written in compact form as $X=[\underline{x},\overline{x}]^n$. Given $x \in \mathbb{R}^n$ and $\varepsilon \geq 0$, the ball with center $x$ and radius $\varepsilon$ is $\ball_{\varepsilon}(x)=\{z \in \mathbb{R}^n \mid \|z-x\|_\infty \leq \varepsilon$\}. 
%%%%%%------------------- Subsec dt Sys ------------------%%%%
\subsection{Interconnected Systems}
In this work, our focus is on the interconnection of discrete-time systems, treating the interconnection effect as a disturbance. Consider a network consisting of $L \in \mathbb{N}_{\geq 1}$ subsystems $\Sigma_i$, $i \in \mathbb{N}_L$. For each $i \in \mathbb{N}_L$, the set of \emph{in-neighbors} of $\Sigma_i$ is denoted by $\mathcal{L}_i \subseteq \mathbb{N}_L \backslash\{i\}$, \emph{i.e.}, the set of subsystems $\Sigma_{\ell}$, $\ell \in \mathcal{L}_i$, directly influencing $\Sigma_i$. 

\begin{definition}[{Discrete-time subsystem}] A discrete-time subsystem $\Sigma_i$ is a tuple $\Sigma_i=$ $\left(X_i, W_i, f_i\right)$, where $X_i \subseteq \mathbb{R}^{n_i}$ and {${W}_i \subseteq \mathbb{R}^{n_i}$} 
are the state and internal input spaces, respectively. $f_i: \mathbb{R}^{n_i} \rightarrow \mathbb{R}^{n_i}$ is called the transition function. 
\end{definition}\label{def dt subsys}
The evolution of the state of the $\Sigma_i$ is given by
\begin{equation}
x_i(k+1)=f_i(x_i(k)) + w_i(k),\quad k \in \mathbb{N},
 \label{eq: dt subsys}
\end{equation}
where $x_i \in X_i$, and
\begin{equation}
\label{eq:disturbance}
    w_i(k) = \sum_{ \ell \in \mathcal{L}_i} {C}_{i,\ell} x_{\ell}(k)\in W_i, 
\end{equation}
with 
$C_{i, \ell}$ indicates the interconnection weight between subsystems having appropriate dimensions. 

A discrete-time interconnected system is introduced as follows.
\begin{definition}[Discrete-time interconnected system]
An interconnected system denoted by $\mathcal{I}\left(\Sigma_1, \ldots, \Sigma_L\right)$ is a tuple $\Sigma=(X, f)$, where $X=\prod_{i \in \mathbb{N}_L} X_i$ is the state space. $f: \mathbb{R}^n \rightarrow \mathbb{R}^n$ is the transition function defined as : $f({x}(k))=$ $\left[f_1\left({x}_1(k)\right)+{w}_1(k) ; \ldots ; f_L\left({x}_L(k)\right)+{w}_L(k)\right]$, where $n=\sum_{i \in \mathbb{N}_L} n_i$, and ${x}=\left[{x}_1 ; \ldots ; {x}_L\right]$. 
The evolution of the state of $\Sigma$ for all $k \in \mathbb{N}$ is given as
\begin{equation}
x(k+1)=f(x(k)).
 \label{eq: dt interconnected sys}
\end{equation}
\end{definition}
%%%%%%------------------- Subsec LTL Spec ------------------%%%%
\subsection{Class of Specifications}
Consider \( \Sigma_i \) as in \eqref{eq: dt subsys}. A specification \( \psi_i \subset X_i^{N+1} \) defines a set of admissible state trajectories over a finite horizon of length \( N \in \mathbb{N} \). This represents the desired system behavior within that time frame. This type of specification is highly expressive and includes properties such as safety, reachability, and more generally, LTL specifications over finite traces, referred to as LTL\(_F\) \cite{zhu2017symbolic}. For instance, the finite-horizon safety specification of a set \( \Gamma_i \subseteq X_i \) over the interval from time step \( 0 \) to \( N \) is denoted by {\( \psi_i = \square^{N} \Gamma_i \)}. 
This property can be expressed as $\psi_i= \Gamma_i^{N+1} \subseteq X^{N+1}_i$.
Likewise, the exact-time reachability specification of a set \( \Gamma_i \subseteq X_i \) at time step \( N \), denoted in LTL\(_F\) as \( \psi_i = \nex^N \Gamma_i \), can also be expressed as
\begin{equation*}
    \psi_i = X^N_i \times \Gamma_i  \subseteq X^{N+1}_i.
\end{equation*}

The finite-horizon reachability for a set \( \Gamma_i \subseteq X_i \) 
within the time interval from \( 0 \) to \( N \) is denoted by \( \psi_i = \eventually^{N} \Gamma_i \), which is 
\begin{equation}
    \psi_i= \bigcup\limits_{j=0}^{N}X^{j}_i\times \Gamma_i \times  X^{N-{j}}_i \subseteq X^{N+1}_i.
\end{equation}
%%%%%%------------------- Subsec LTL Spec ------------------%%%%
\subsection{Resilience for Temporal Specifications}
\label{sec: Resilience LTL}
Consider the subsystem \( \Sigma_i \) defined in \eqref{eq: dt subsys}. We denote \( \xi_i(\bar{x}_i, w_i) \) as the state trajectory of subsystem \( \Sigma_i \) over a finite time horizon of length \( N \in \mathbb{N} \).
This trajectory starts from an initial condition \( \bar{x}_i \in X_i \) and is influenced by an internal input sequence \( w_i = (w_i(0), w_i(1), \ldots, w_i(N-1)) \in W_i^N \), which represents the effects of neighboring subsystems and will be treated as a disturbance in this work. Additionally, we use  \(\xi(\bar{x}_i, W_i) \) to represent the set of all such trajectories that start from \( \bar{x}_i \in X_i \) under every admissible internal input sequence \( w_i \in W_i^N \). {For $W_i=\ball_{\varepsilon_i}(0)$,} this collection of trajectories is formally defined as 
\begin{align}
\label{eqn:reach}
    \xi_i(\bar{x}_i, \varepsilon_i) &:= \{(x_i(0),x_i(1), \dots, x_i(N)) \mid x_i(0) = \bar{x}_i, \nonumber\\
    &x_i(k+1) = f(x_i(k) ) + w_i(k), w_i(k) \in W_i\}.
\end{align}
Note that $\xi_i(\bar{x}_i,0)$ contains only the disturbance-free trajectory of the system, known as the nominal trajectory. This indicates that neighboring subsystems do not impact this subsystem, and $w_i = 0$. Similarly, given a set $B \subseteq X_i$, we define the set of trajectories $\xi_i(B,\varepsilon_i)=\bigcup\limits_{\bar{x}_i \in B} \xi_i(\bar{x}_i,\varepsilon_i).$ The system $\Sigma_i$ starting from  \( \bar{x}_i \in X_i \) is said to satisfy the specification $\psi_i \subseteq X_i^{N+1}$ denoted by $\Sigma_i \vDash \psi_i $ if $\xi_i( \bar{x}_i,W_i) \subseteq \psi_i$. 

We adopt the concept of resilience as outlined in \cite{saoud2024temporal} to define the resilience metric as follows, in which it is assumed that the disturbance is located within a ball-shaped set as \( W_i = \Omega_{\varepsilon_i} (0)\). 
\begin{definition}[Resilience metric]
\label{def:resilience}
Consider the subsystem $\Sigma_i$ in \eqref{eq: dt subsys}, a specification $\psi_i$ and a point \( \bar{x}_i \in X_i \). We define the {\it resilience} of the subsystem $\Sigma_i$ with respect to the initial condition of \( \bar{x}_i\) and the specification $\psi_i$ as a function $g_{\psi_i}:X_i\rightarrow \reals_{\ge 0}\cup\{+\infty\}$ with
\begin{equation}
\label{eq:quan} g_{\psi_i}(\bar{x}_i)\!\!=\!\!\begin{cases}
\sup\left\{\varepsilon_i\ge 0\,|\,\xi_i(\bar{x}_i,\varepsilon_i)\vDash \psi_i\right\},&\text{if } \xi_i(\bar{x}_i,0)\vDash\psi_i,\\
0, &\text{if } \xi_i(\bar{x}_i,0)\nvDash\psi_i,
\end{cases}
\end{equation}
where $\xi_i(\bar{x}_i,\varepsilon_i)\vDash \psi_i$ indicates that all trajectories in $\xi_i(\bar{x}_i,\varepsilon_i)$ satisfy the specification.
\end{definition}

According to Proposition 3.1 in \cite{saoud2024temporal}, the following structural properties hold:
\begin{itemize}
    \item[(i)] For any specification $\psi_i = \psi_i^1\wedge\psi_i^2$, we have that $g_{\psi_i}(\bar{x}_i) = \min\{g_{\psi_i^1}(\bar{x}_i),g_{\psi_i^2}(\bar{x}_i)\}$, for all \( \bar{x}_i \in X_i \).
    \item[(ii)] For any specification $\psi_i = \psi_i^1 \vee \psi_i^2$, we have that $g_{\psi_i}(\bar{x}_i) \geq \max\{g_{\psi_i^1}(\bar{x}_i),g_{\psi_i^2}(\bar{x}_i)\}$, for all \( \bar{x}_i \in X_i \). 
    \item[(iii)] For $B_i \subset X_i$, ${g_{\psi_i}(B_i)= \inf_{\bar{x}_i \in B_i}g_{\psi_i}(\bar{x}_i)}$.
\end{itemize}
Note that property (iii) implies that, for a set $B_i \subseteq X_i$, the resilience metric over the set $B_i$ and specification $\psi_i$ can be defined as follows
\begin{align}\nonumber
&g_{\psi_i}(B_i) = \\\label{eq res metric A}&\!\!
\begin{cases}
\sup\left\{\varepsilon_i\ge 0|\xi_i(B_i,\varepsilon_i)\vDash\psi_i\right\},\!\!\!\!\!&\text{if }{\forall} \bar{x}_i\!\in\! B_i, \xi_i(\bar{x}_i,0)\!\vDash\!\psi_i,\\
0,&\text{if }{\exists} \bar{x}_i\!\in\! B_i, \xi_i(\bar{x}_i,0)\!\nvDash\!\psi_i. 
\end{cases} 
\end{align}
%%%%%%------------------- Assume Guarantee ------------------%%%%
\subsection{Compositional Reasoning via AGCs}
\label{sec: AG Reasoning}
To ensure that temporal logic specifications are met in interconnected systems, we utilize the framework of AGCs. AG reasoning supports a compositional verification approach, which establishes the correctness of the overall system by verifying individual components within specific interaction constraints. The formal definition of AGCs for an interconnected system is provided below.
\begin{definition}[AGCs]
\label{def agcs}
Consider a subsystem $\Sigma_i=(X_i,W_i,f_i)$. An AGC for $\Sigma_i$ is a tuple $\mathcal{C}_i=\left(\mathcal{A}_i, \mathcal{G}_i\right)$ where
\begin{itemize}
    \item $\mathcal{A}_i \subseteq W_i^{N+1}$ is the assumption specification,
    \item $\mathcal{G}_i \subseteq X_i^{N+1}$ is the guarantee specification.
\end{itemize}

We say that $\Sigma_i$ satisfies $\mathcal{C}_i$, denoted by $\Sigma_i \models_{\mathfrak{c}} \mathcal{C}_i$, if for all trajectories $(x_i,w_i): \mathbb N_N \rightarrow X_i \times W_i $,
$w_i\models \mathcal{A}_i$ implies $x_i\models \mathcal{G}_i$. 
\end{definition}

\subsection{Problem Statement}
\begin{resp1}
\begin{problem}
\label{prblm}
    Given the interconnected system $\Sigma$ in \eqref{eq: dt interconnected sys} with individual subsystems $\Sigma_i$ in \eqref{eq: dt subsys} and local specifications $\psi_i\subseteq X^{N+1}$, compute 
    a collection of AGCs \(\mathcal{C}_i = (\mathcal{A}_i, \mathcal{G}_i)\), such that \(\Sigma_i \vDash_{\mathfrak{c}}\mathcal{C}_i\) and \(\Sigma \models \psi_1\wedge \psi_2\wedge \ldots\wedge
\psi_n\).
\end{problem}
\end{resp1}

The term \(w_i(k)\) represents the internal input of subsystem \(\Sigma_i\), capturing the influence of other subsystems on \(\Sigma_i\).
To provide a solution for the above problem, we treat the internal inputs of the subsystem \(\Sigma_i\)  as a disturbance \(w_i(k)\) and design the contracts based on the resilience of the subsystems to such disturbances, as described in the next section.
%%%%%%%%%%%%%%%%%%%%%%%%%%%%%%%%%%%%%%%%%%%%%%%%%%%%%%%%%%%%%%%%%%%%%%%%%%%%%%%%
\section{Resilience-based AG Reasoning} \label{sec AG reasoning}
\subsection{Two Interconnected Subsystems}
We now explain the resilience-based AG procedure for two interconnected subsystems, \emph{i.e.} $L=2$. Each subsystem $\Sigma_i$ is associated with a local specification $\psi_i \subseteq X_i^{N+1}$, while the coupling term, \emph{i.e.}, $C_{1,2}x_2$ in $\Sigma_1$ or $C_{2,1}x_1$ in $\Sigma_2$, is treated as a structured disturbance. As provided in Algorithm~\ref{Alg: AG },
at iteration $j$, subsystem $\Sigma_1$ first computes the maximum disturbance radius $\varepsilon_1^j$ for which its current specification $\psi_1^j$ is satisfied. This defines an assumption set $\mathcal{A}_1^j = \square^N \Omega_{\varepsilon_1^j}(0)$ 
on the internal input sequence of $\Sigma_1$. Using the interconnection relation~\eqref{eq:disturbance}, this assumption is translated into a finite-horizon safety specification as
\(\phi_2^j := \{x_2 \in X_2^N \mid C_{1,2}x_2 \vDash \mathcal{A}_1^j\}\),
which refines the specification of subsystem $\Sigma_2$ to $\psi_2^{j+1} = \psi_2^j \wedge \phi_2^j$.  

Next, subsystem $\Sigma_2$ computes its resilience value $\varepsilon_2^j$ with respect to the refined specification $\psi_2^{j+1}$, leading to the assumption set $\mathcal{A}_2^j = \square^N \Omega_{\varepsilon_2^j}(0)$. This is again translated through~\eqref{eq:disturbance} into a finite-horizon safety specification as
\(\phi_1^j := \{x_1 \in X_1^N \mid C_{2,1}x_1 \vDash \mathcal{A}_2^j\}\),
which updates the specification of subsystem $\Sigma_1$ to $\psi_1^{j+1} = \psi_1^j \wedge \phi_1^j$. In essence, each iteration involves two updates: first, we refine the specification of \(\Sigma_2\) based on the resilience of \(\Sigma_1\), and then we refine the specification of \(\Sigma_1\) based on the resilience of \(\Sigma_2\). This iterative process continues until no further refinements are made, meaning that \(\psi_1^{j+1} = \psi_1^j\) and \(\psi_2^{j+1} = \psi_2^j\). Once convergence is achieved, the resulting assumptions \(\mathcal{A}_1^{j+1}\) and \(\mathcal{A}_2^{j+1}\), along with the guarantees \(\mathcal{G}_1 = \psi_1^{j+1}\) and \(\mathcal{G}_2 =\psi_2^{j+1}\), create feasible AGCs that ensure mutual compatibility.   
%%%%%%%%%%%%%%%%%%%%%% Algorithm %%%%%%%%%%%%%%%%%%%%%%
\RestyleAlgo{ruled}
\IncMargin{0.1em}
\begin{algorithm}
\hspace{3mm} 
\begin{minipage}{\dimexpr\linewidth-3mm}
\SetAlgoLined
\SetKwInOut{Input}{Input}
\SetKwInOut{Output}{Output}
\SetKwFunction{GenerateRandomNumber}{GenerateRandomNumber}
\SetKwFunction{Wait}{Wait}
\SetKwFunction{Break}{\textbf{Break}}
\caption{Resilience-based Computation of AGCs for two interconnected subsystems} 
\label{Alg: AG }
\setcounter{AlgoLine}{0}
\Input{Subsystems $\Sigma_1$ and $\Sigma_2$, specifications $\psi_1$ and $\psi_2$.}
\BlankLine
\SetAlgoLined

$\psi_1^1 := \psi_1$, and $\psi_2^1 := \psi_2$ 

\For{$j \in \mathbb{N}$}
{
        Compute the resilience metric $\varepsilon_1^j$ using \eqref{eq res metric A} such that $\Sigma_1 \vDash \psi_1^j$ 

        Define the assumption for the internal input sequence as $\mathcal{A}_1^j =  \always^{N}\Omega_{\varepsilon_1^j}(0) \subseteq W_1^N$ 

        Translate $\mathcal{A}_1^j$ into a finite-horizon safety specification as $\phi_2^j := \{x_2\in X_2^{N}\,|\, C_{1,2}x_2\vDash \mathcal{A}_1^j \}$ for $\Sigma_2$ using \eqref{eq:disturbance}
        
        Define the new specification for $\Sigma_2$ as $\psi_2^{j+1} = \psi_2^j \wedge \phi_2^j$ \;

        Compute the resilience metric $\varepsilon_2^j$ using \eqref{eq res metric A} such that $\Sigma_2 \vDash \psi_2^{j+1}$

        Define the assumption for the internal input sequence as $\mathcal{A}_2^j =  \always^{N}\Omega_{\varepsilon_2^j}(0) \subseteq W_2^N$ 

        Translate $\mathcal{A}_2^j$ into a finite-horizon safety specification as $\phi_1^j := \{x_1\in X_1^{N}\,|\, C_{2,1}x_1\vDash \mathcal{A}_2^j \}$ for $\Sigma_1$ by \eqref{eq:disturbance}

        Define new specification for $\Sigma_1$ as $\psi_1^{j+1} = \psi_1^j \wedge \phi_1^j$ \;

        \If{$\psi_1^{j+1} = \psi_1^{j}$ and $\psi_2^{j+1} = \psi_2^{j}$}{
            \Break
        }
}

$\mathcal{A}_1 = \mathcal{A}_1^{j+1}$, $\mathcal{A}_2 = \mathcal{A}_2^{j+1}$,
$\mathcal{G}_1 = \psi_1^{j+1}$, and
$\mathcal{G}_2 = \psi_2^{j+1}$

\Output{Contracts $\mathcal{C}_1 = (\mathcal{A}_1, \mathcal{G}_1)$ and $\mathcal{C}_2 = (\mathcal{A}_2, \mathcal{G}_2)$.}
\BlankLine
\end{minipage}
\end{algorithm}
%%%%%%%%%%%%%%%%%%%%%%%%%%%%%%
The following theorems state properties of the solutions constructed by Algorithm~\ref{Alg: AG }.

\begin{theorem}[Correctness]
 \label{thm  converge}
    If Algorithm \ref{Alg: AG } terminates, it generates contracts that are solutions for Problem~\ref{prblm} for the case of two subsystems.
\end{theorem}

\begin{theorem}[Monotonicity]
 \label{thm monotonic}
    The sequence of guarantees $\{\psi_1^j,j\in\mathbb N\}$ and $\{\psi_2^j,j\in\mathbb N\}$  computed in Algorithm \ref{Alg: AG } are monotonically non-increasing with respect to set inclusion.
\end{theorem}

\subsection{$L$ Interconnected Subsystems}
We generalize Algorithm~\ref{Alg: AG } to address \(L\) interconnected subsystems as presented in Algorithm~\ref{Alg: AG FS 2}. Each subsystem \(\Sigma_i\) is affected by its neighboring subsystems \(\Sigma_\ell\), \(\ell \in \mathcal{L}_i\). These influences are modeled as structured disturbances \(w_i\) as in \eqref{eq:disturbance}. At each iteration $j$, the algorithm computes the resilience \(\varepsilon_i^j\) for each subsystem \(\Sigma_i\), determining the maximum disturbance it can withstand while still meeting its local specification \(\psi_i^j\). The resulting disturbance ball, \(\Omega_{\varepsilon_i^j}(0)\), is then decomposed into component-wise bounds on each neighbor’s contribution, utilizing a set of non-negative weights
$\Lambda^i:=\{\lambda_\ell^i, {\ell} \in \mathcal{L}_i\mid \lambda_{\ell}^i\ge 0, \sum_{\ell \in \mathcal{L}_i} \lambda_{\ell}^i  =1\}$.
These weights quantify the relative influence of each neighbor \(\Sigma_\ell\) on \(\Sigma_i\) and can be selected based on prior knowledge or heuristics on the dynamics of the subsystems. By assigning a larger $\lambda_\ell^i$, subsystem $\Sigma_\ell$ is expected to handle a greater share of the disturbance, whereas a smaller $\lambda_\ell^i$ limits its allocated disturbance. Therefore, if a particular neighbor $\Sigma_\ell$ is known to be sensitive to disturbances, $\lambda_\ell^i$ can be set higher relative to other subsystems to mitigate its exposure. 
Each component disturbance bound \(\Omega_{\lambda_\ell \varepsilon_i^j}(0)\) is then translated into a finite-horizon safety specification $\phi_i^j := \{x_i\in X_i^{N}\,|\, C_{\ell,i}x_i\in \mathcal{A}_\ell^j \}$  for the neighbor \(\Sigma_\ell\). This specification is then added to the current local specification \(\psi_\ell^j\) to produce \(\psi_\ell^{j+1}\). This iterative process continues across all subsystems until convergence is reached. The final assumption sets \(\mathcal{A}_i\) define feasible AGCs \(\mathcal{C}_i = (\mathcal{A}_i, \mathcal{G}_i)\) for each subsystem.
%%%%%%%%%%%%%%% Algorithm Finite Safety m sub %%%%%%%%%%%%%%%
\RestyleAlgo{ruled}
\IncMargin{0.1em}
\begin{algorithm}
\hspace{3mm} 
\begin{minipage}{\dimexpr\linewidth-3mm}
\SetAlFnt{\small}
% \scriptsize
\SetAlgoLined
\SetKwInOut{Input}{Input}
\SetKwInOut{Output}{Output}
\SetKwFunction{GenerateRandomNumber}{GenerateRandomNumber}
\SetKwFunction{Wait}{Wait}
\SetKwFunction{Break}{\textbf{Break}}
\caption{Resilience-based Computation of AGCs for $L$ interconnected subsystems} 
\label{Alg: AG FS 2}
\setcounter{AlgoLine}{0}
\Input{Subsystems $\Sigma_i$ and specifications $\psi_i$ for $ i \in \mathbb{N}_L$.}
\BlankLine
\SetAlgoLined
$\psi_i^1 := \psi_i$ for $ i \in \mathbb{N}_L$

\For{$j \in \mathbb{N}$}
{
Compute the resilience metric $\varepsilon_i^j$ using \eqref{eq res metric A} such that $\Sigma_i \vDash \psi_i^j, \forall i \in \mathbb{N}_L$

Define the assumptions for the internal input sequence of each subsystem as $\mathcal{A}_i^j=\always^N\Omega_{\varepsilon_i^j}(0) \subseteq W_i^N, \forall i \in \mathbb{N}_L$

Define $\Lambda^i:=\{\lambda_\ell^i, {\ell} \in \mathcal{L}_i\mid \lambda_{\ell}^i\ge 0, \sum_{\ell \in \mathcal{L}_i} \lambda_{\ell}^i  =1\}$ based on the knowledge of the subsystems

Decompose assumption $\mathcal{A}_i^j$ based on the sequence of the disturbance in \eqref{eq:disturbance} and $\lambda^i_\ell$ to $\mathcal{A}_{i,\ell}^j = \Omega_{\lambda_\ell^i\varepsilon_i^j}(0)$ 

Translate $\mathcal{A}_{i,\ell}^j$ into a finite-horizon specifications on each $x_\ell$ for $\Sigma_\ell$ for all $\ell \in \mathcal{L}_i$ as  $\phi^j_i = \{x_i \in X_i^N | C_{i,\ell}x_\ell \in {\mathcal{A}_{i,\ell}^j}, \forall i \in \mathbb{N}_L, \forall \ell \in \mathcal{L}_i$\} by \eqref{eq:disturbance}

Define new specification for each subsystem $\Sigma_i$ as $\psi_i^{j+1} = \psi_i^j \wedge (\bigwedge_{\ell \in \mathcal{L}_i}\phi_{i,\ell}^j), \forall i \in \mathbb{N}_L$

\If{$\psi_i^{j+1} = \psi_i^{j}$, $\forall i \in \mathbb{N}_L$}{
            \Break
        }
}

$\mathcal{A}_i = \mathcal{A}_i^{j+1}$ and
$\mathcal{G}_i = \psi_i^{j+1}$, $\forall i \in \mathbb{N}_L$

\Output{Contracts $\mathcal{C}_i=(\mathcal{A}_i,\mathcal{G}_i)$, $\forall i \in \mathbb{N}_L$}
\BlankLine
\end{minipage}
\end{algorithm}
%%%%%%%%%%%%%%%%%%%%%%%%%%%%%%
The following theorem is provided to demonstrate the correctness of Algorithm \ref{Alg: AG FS 2}.
\begin{theorem}
\label{thm convergence l subsys}
If Algorithm~\ref{Alg: AG FS 2} terminates, it generates contracts that are solutions for Problem~\ref{prblm} for the case of $L$ subsystems.
\end{theorem}

\begin{remark}
The assumption refinement process in Algorithm~\ref{Alg: AG FS 2} uses weights \( \lambda_\ell^i \), which determine the influence of each neighboring subsystem on the assumption updates. These weights can be adjusted based on the relative importance of each subsystem, providing greater flexibility in contract synthesis. This is at the cost of losing maximality for $L > 2$.
For two subsystems, the assumption refinement process is directly linked to the computation of the resilience metric. However, when multiple subsystems are involved, the weighted influence introduces additional constraints, which may lead to more conservative sets of assumptions. Consequently, the computed sets of assumptions for three or more interconnected subsystems may not be the largest possible sets that ensure the satisfaction of the specified requirements.
\end{remark}
\begin{remark}
We have modeled internal inputs as linear additive combinations of neighboring states as in \eqref{eq:disturbance}, which provides a clear formulation for the resilience-based AG reasoning framework. More generally, the interconnection can be nonlinear, \emph{i.e.} 
$w_i(k) = f_i(x_1(k),\dots,x_\ell(k))$, and if $f_i$ is bijective and invertible, the framework extends naturally. The linear additive case thus serves as a simplified but representative case.
\end{remark}
\section{Computation of Feasible AGCs} \label{sec com lin}
In this section, we demonstrate how AGCs can be computed for linear and nonlinear systems using the concept of resilience metric and the algorithms of the previous section. The numerical examples of this section are implemented in MATLAB using optimization toolboxes $\mathsf{CVX}$ and $\mathsf{MOSEK}$ \cite{grant2014cvx, aps2019mosek}.
All the simulations of this paper are performed on a $\mathsf{MacBook Pro}$ with the $\mathsf{M2 Pro}$ chip and 16GB of memory.

\subsection{Linear Systems}
Consider the special case where the interconnected subsystems defined in \eqref{eq: dt subsys}--\eqref{eq: dt interconnected sys} are linear. A linear subsystem is a special case of the general nonlinear system defined in equation~\eqref{eq: dt subsys}, given by
\begin{equation}
\Sigma_i: x_i(k+1) = A_i x_i(k) + w_i(k),\quad \forall k \in \mathbb{N}, \label{eq linear sigma}
\end{equation}
where \( A_i \) is a constant system matrix, and \( w_i(k) \) denotes the interconnection effect, which is considered a disturbance within the resilience concept as previously defined. 
Resilience metrics for linear subsystems can be calculated based on the findings presented in \cite{saoud2024temporal}, with different formulations applicable to various types of specifications. Below, we summarize these cases and provide numerical examples to illustrate them. These resilience computations are essential components used in Step~3 and Step~7 of Algorithm~\ref{Alg: AG }, as well as in Step~3 of Algorithm~\ref{Alg: AG FS 2}.

\subsubsection{Finite-Horizon Safety} \label{sec FS lin}
For the finite-horizon safety specification $\psi_i = \square^N \Gamma_i$, where $\Gamma_i=\{x_i\in X_i\,|\, G_i x_i\le H_i\}$, for some $N \in \mathbb{N}$, resilience is computed using \cite[Theorem~4.2]{saoud2024temporal}. Example~\ref{ex 1} illustrates the implementation on three interconnected subsystems.
\begin{example} \label{ex 1}
    Consider three interconnected linear subsystems as in \eqref{eq linear sigma} with
    $A_1 = \begin{bmatrix}
        0.1 & -0.5 \\
        -0.1 & 0.7
    \end{bmatrix}$, $A_2= \begin{bmatrix}
        0.6 & 0 \\
        -0.1 &  0.6
    \end{bmatrix}$, $A_3 = \begin{bmatrix}
        -0.5 & 0 \\
        0 & -0.5
    \end{bmatrix}$, $C_{1,2} = C_{1,3}= \begin{bmatrix}
        0.008 & 0.001 \\
        -0.006 & -0.009
    \end{bmatrix}$, $C_{2,1}= \begin{bmatrix}
        -0.001 & 0.002 \\
        0.001 & -0.001
    \end{bmatrix}$, $C_{2,3}= \begin{bmatrix}
        -0.005 & 0.002 \\
        0.001 & -0.005
    \end{bmatrix}$, and $C_{3,1} = \begin{bmatrix}
        0.001 & 0 \\
        0 & 0.001
    \end{bmatrix}$.
    Three polytopes are defined as $\Gamma_1 = \{x_1 \in X_1| G_1x_1 \leq H_1\}$,  $\Gamma_2 = \{x_2 \in X_2| G_2x_2 \leq H_2\}$ and $\Gamma_3 = \{x_3 \in X_3| G_3x_3 \leq H_3\}$ with $G_1 = \begin{bmatrix}
            1 & 0.1 & -1 & 0.1 \\
            0.1 & 1 & 0.1 & -1
        \end{bmatrix}^\top$, $G_2 = \begin{bmatrix}
            2 & 0 & -1 & 0.2 \\
            0 & 2 & 0.2 & -1
        \end{bmatrix}^\top$, $G_3 = \begin{bmatrix}
            1 & 0 & -1 & 0 \\
            0 & 1 & 0 & -1
        \end{bmatrix}^\top$, $H_1=H_2=\begin{bmatrix}
            10&10&14&14
        \end{bmatrix}^\top$, and $H_3=\begin{bmatrix}
            10&10&10&10
        \end{bmatrix}^\top$.
The objective is to consider the finite-horizon safety specifications defined by $\psi_1 = \square^3\Gamma_1$, $\psi_2 = \square^3\Gamma_2$, and $\psi_3 = \square^3\Gamma_3$ for $\Sigma_1$, $\Sigma_2$ and $\Sigma_3$ , respectively. Algorithm~\ref{Alg: AG FS 2} on this example terminates after 3 iterations and provides the contracts.
We then simulate the system using 100 random initial conditions from the assumptions and observe that all the trajectories remain within the safe set (see Fig.~\ref{fig: EX FS}).
\begin{figure*}
    \centering
\includegraphics[width=0.32\linewidth]{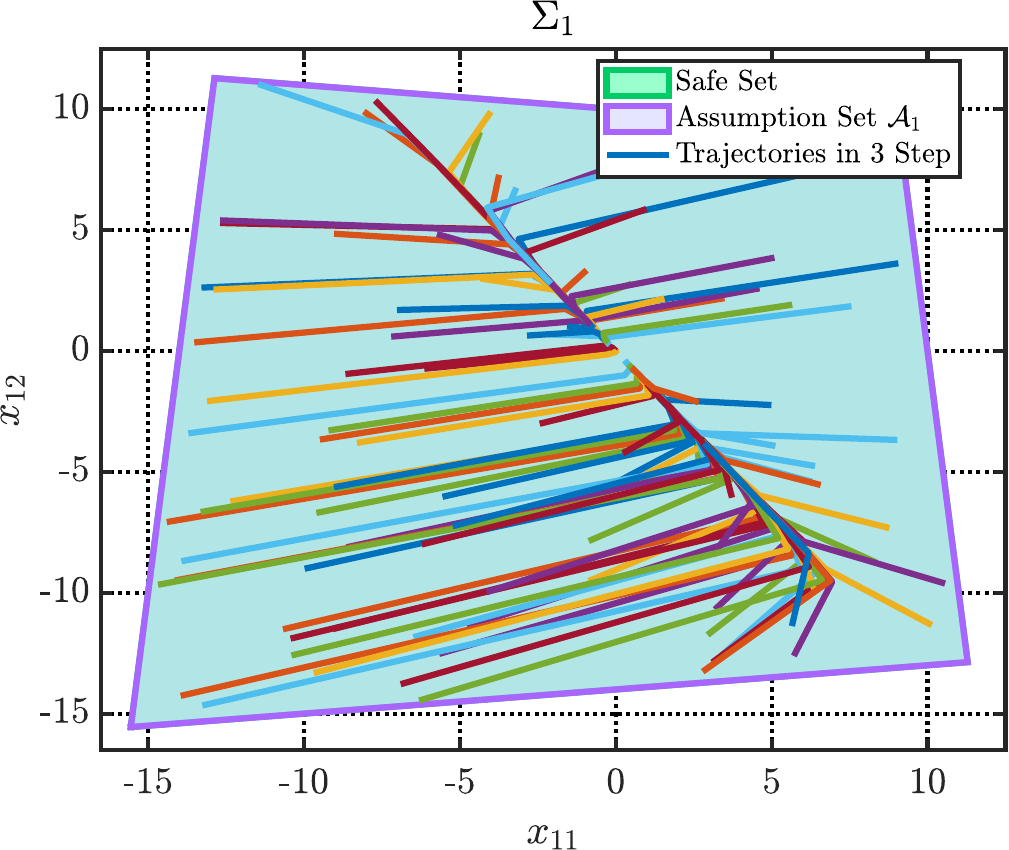}
    % \vspace{1mm}
\includegraphics[width=0.32\linewidth]{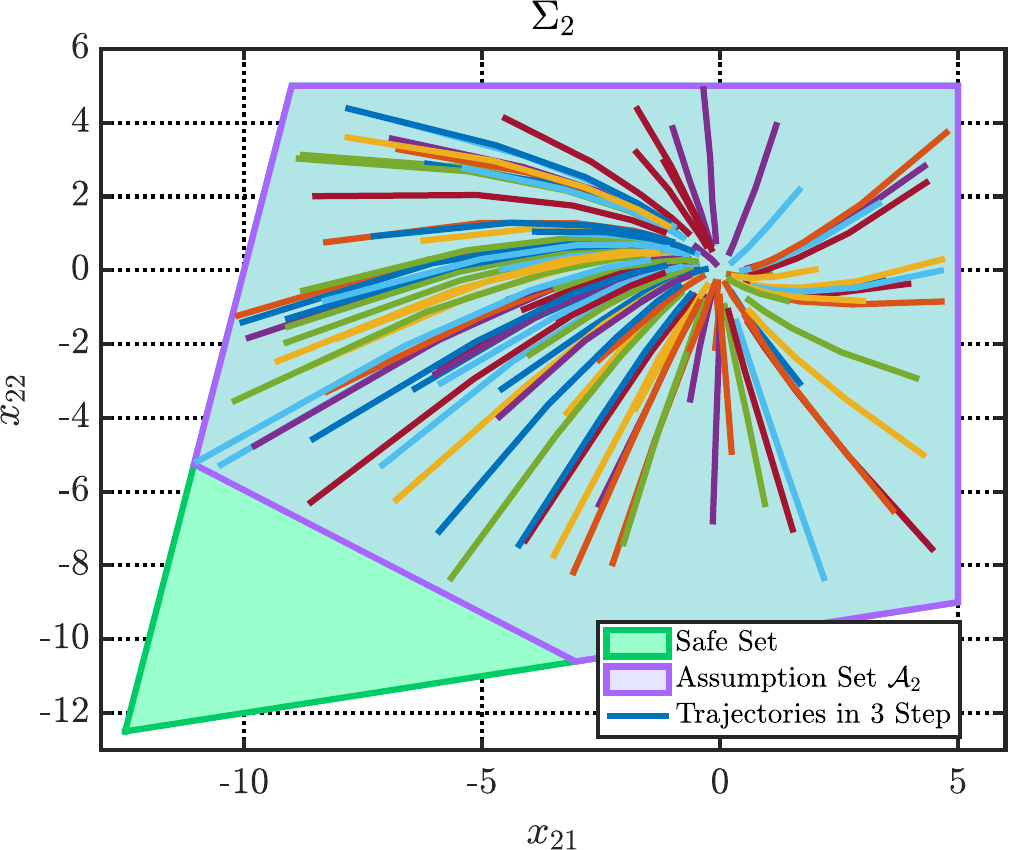}
\includegraphics[width=0.32\linewidth]{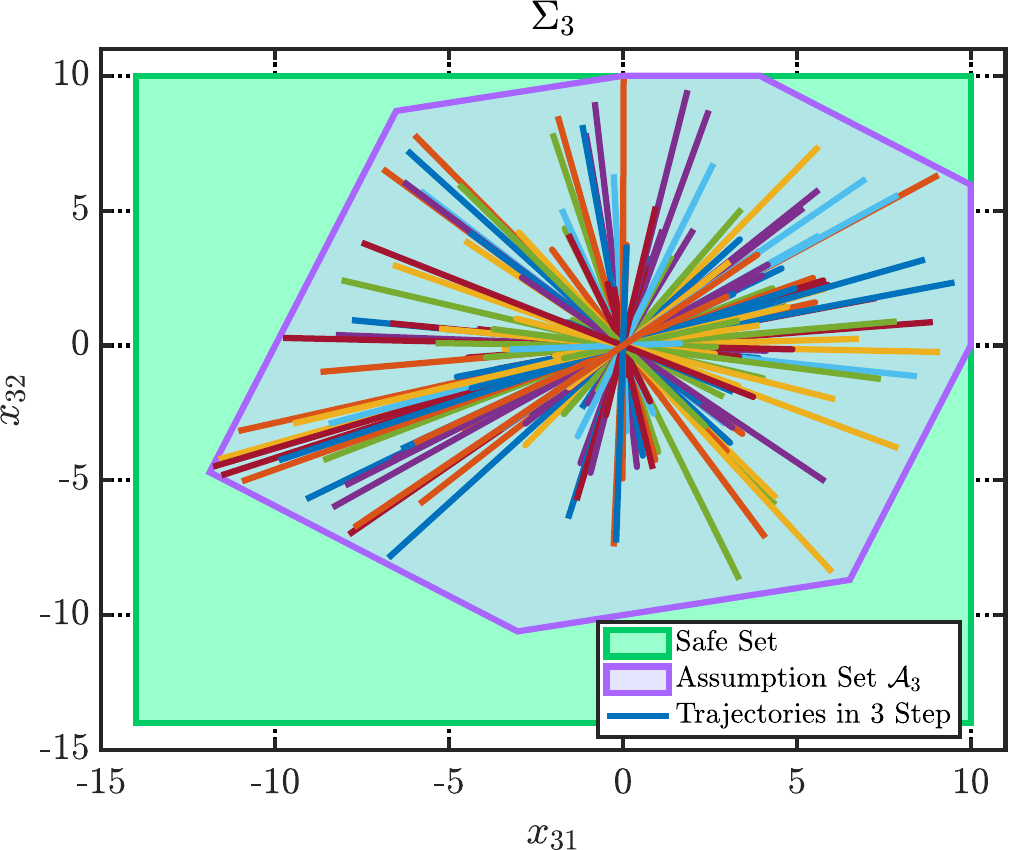}
    \caption{Assumption sets calculated by Algorithm~\ref{Alg: AG FS 2}, and 100 realizations of trajectories started from assumption sets while remaining in the safe set.}
    \label{fig: EX FS}
\end{figure*}
\end{example}
\subsubsection{Exact-Time Reachability}
For the exact-time reachability specification $\psi_i = \nex^N \Gamma_i$, where $\Gamma_i=\{x_i\in X_i\,|\, G_i x_i\le H_i\}$, for some $N \in \mathbb{N}$, the resilience is calculated using \cite[Theorem 4.1]{saoud2024temporal} in Step 3 and Step 7 of Algorithm~\ref{Alg: AG }, as well as in Step 3 of Algorithm~\ref{Alg: AG FS 2}. Example~\ref{ex er}  demonstrates the implementation involving two interconnected subsystems.
\begin{example} \label{ex er}
    Consider two interconnected linear systems as in \eqref{eq linear sigma} with
    $A = \begin{bmatrix}
        0.1 & -1 \\
        -0.5 & -0.2
    \end{bmatrix}$, $C_{12} = \begin{bmatrix}
        0.001 & 0 \\
        0 & 0.001
    \end{bmatrix}$, and $C_{21}=\begin{bmatrix}
        0.0004 & 0 \\
        0 & 0.0004
    \end{bmatrix}$. Two poly-topes are defined as $\Gamma_1 = \{x_1 \in X_1| G_1x_1 \leq H_1\}$ and $\Gamma_2 = \{x_2 \in X_2| G_2x_2 \leq H_2\}$ with $G_1 = G_2 = \begin{bmatrix}
            1 & 0 & -1 & 0 \\
            0 & 1 & 0 & -1
        \end{bmatrix}^\top$, $H_1=\begin{bmatrix}
            10&10&14&14
        \end{bmatrix}^\top$, and $H_2=\begin{bmatrix}
            9&9&9&9
        \end{bmatrix}^\top$.
Consider the exact-time reachability specifications $\psi_1 = \nex^3\Gamma_1$ and $\psi_2 = \nex^3\Gamma_2$ for $\Sigma_1$ and $\Sigma_2$, respectively. 
Algorithm~\ref{Alg: AG } terminates in the third iteration and provides the contracts. 
We simulate the system with 10 random initial conditions, and all the trajectories reach the target set exactly at the third step (see Fig.~\ref{fig: Ex ER}).

\begin{figure}
    % \centering
    \includegraphics[width=0.49\linewidth]{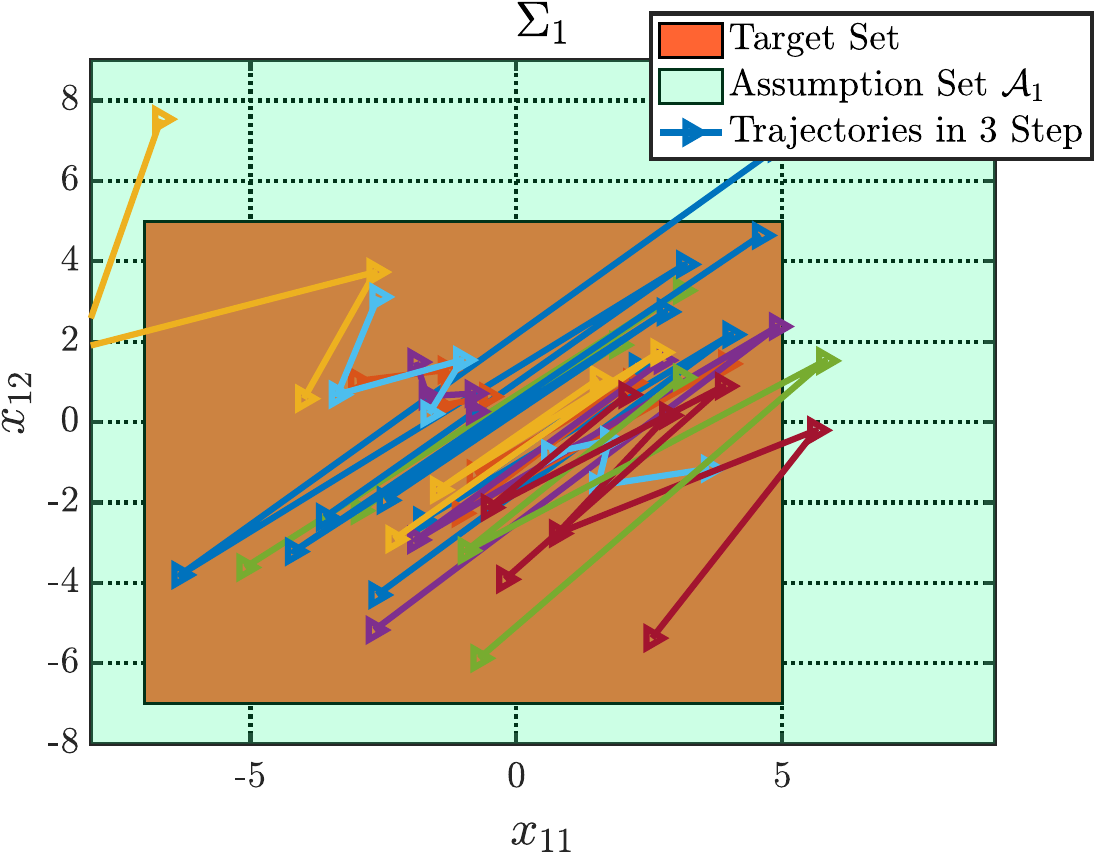}
    \includegraphics[width=0.49\linewidth]{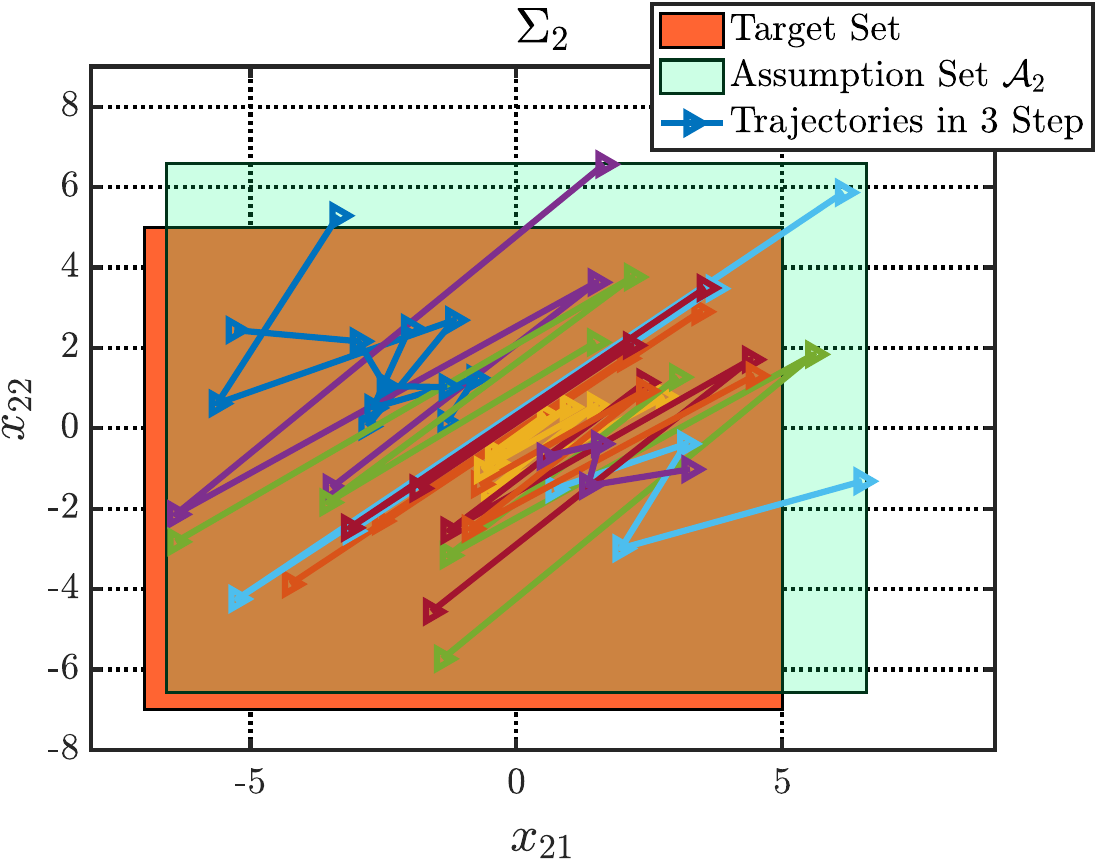}
    \caption{Assumption sets calculated by Algorithm~\ref{Alg: AG }, and 15 trajectories starting from assumption sets and reaching the target set in step 3.}
    \label{fig: Ex ER}
\end{figure}
\end{example}

\subsubsection{Finite-Horizon Reachability}
For the finite-horizon reachability specifications $\psi_i = \lozenge^N \Gamma_i$, where $\Gamma_i=\{x_i\in X_i\,|\, G_i x_i\le H_i\}$, for some $N \in \mathbb{N}$, resilience metric is approximated using scenario optimization as in \cite[Section~\rom{4}.C.2, Theorem 4.3]{saoud2024temporal}. These approximated resilience values are used in Step 3 and Step 7 of Algorithm~\ref{Alg: AG }, as well as in Step 3 of Algorithm~\ref{Alg: AG FS 2}. Example \ref{ex3} presents the implementation of two interconnected subsystems.
\begin{example} \label{ex3} 
    Consider two interconnected linear dynamical systems $\Sigma_1$ and $\Sigma_2$ with the same dynamics as in Example~\ref{ex er}, and two polytopes $\Gamma_1$, and $\Gamma_2 $ as in Example~\ref{ex er}. Consider finite-horizon reachability specifications defined by $\psi_1 = \lozenge^4\Gamma_1$, and $\psi_2 = \lozenge^4\Gamma_2$ as objectives for $\Sigma_1$, and $\Sigma_2$, respectively.
    
    We apply Algorithm~\ref{Alg: AG } and compute the required resilience metric in the first iteration and step 3 of the algorithm using the approximations in 
    \cite[Theorem~4.3]{saoud2024temporal}, which is formulated as a scenario optimization problem (see \cite{esfahani2014performance, garatti2025non, monir2025robust}). 

    We fix the scenario parameters to $\eta = 0.01$ and $\beta = 0.05$. Based on these parameters, we need to take 13 samples to solve the optimization problem. Hence, the resilience metric that we find in the first iteration of the algorithm is with $95\%$ confidence. In the subsequent iterations, the specification will be updated to the finite safety specification, and we use \cite[Theorem~4.3]{saoud2024temporal} for the computations. The algorithm terminates at iteration 3. 
    We then simulate the system using 15 random initial conditions from the assumptions and observe that the trajectories reach the target set eventually at step 4 (see Fig.~\ref{fig: Ex FR}). 
\begin{figure}
    % \centering
    \includegraphics[width=0.49\linewidth]{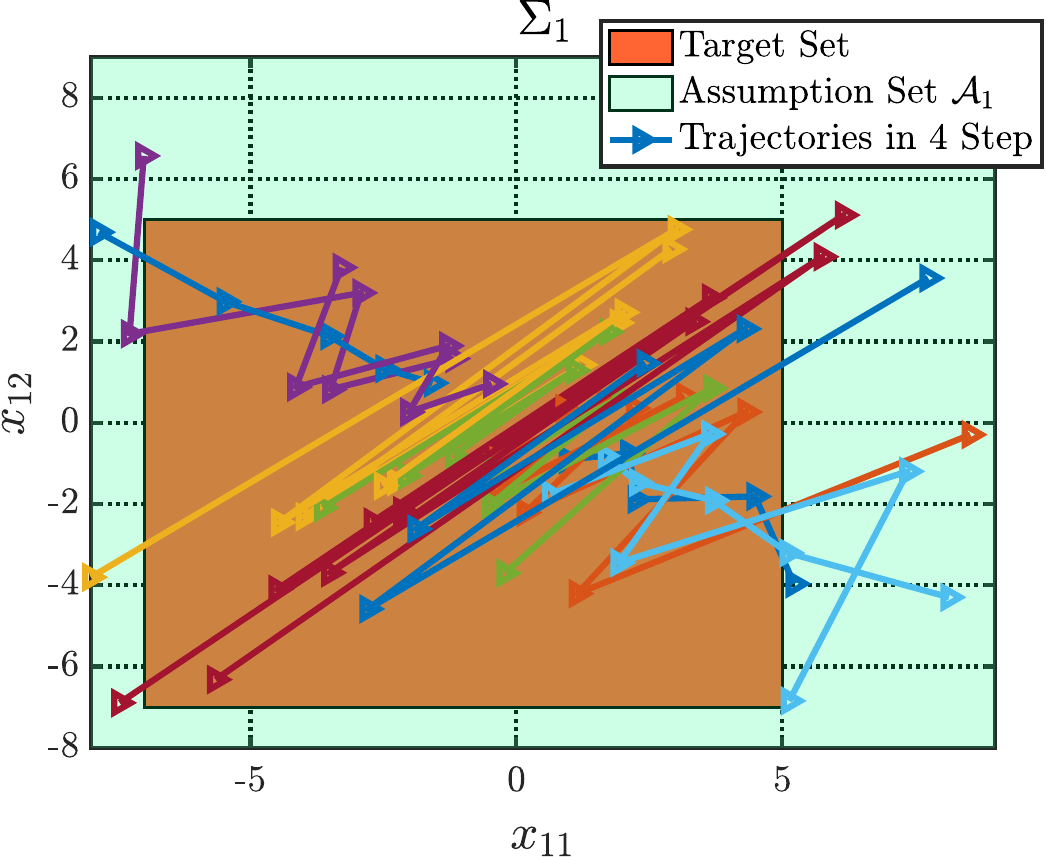}
    \includegraphics[width=0.49\linewidth]{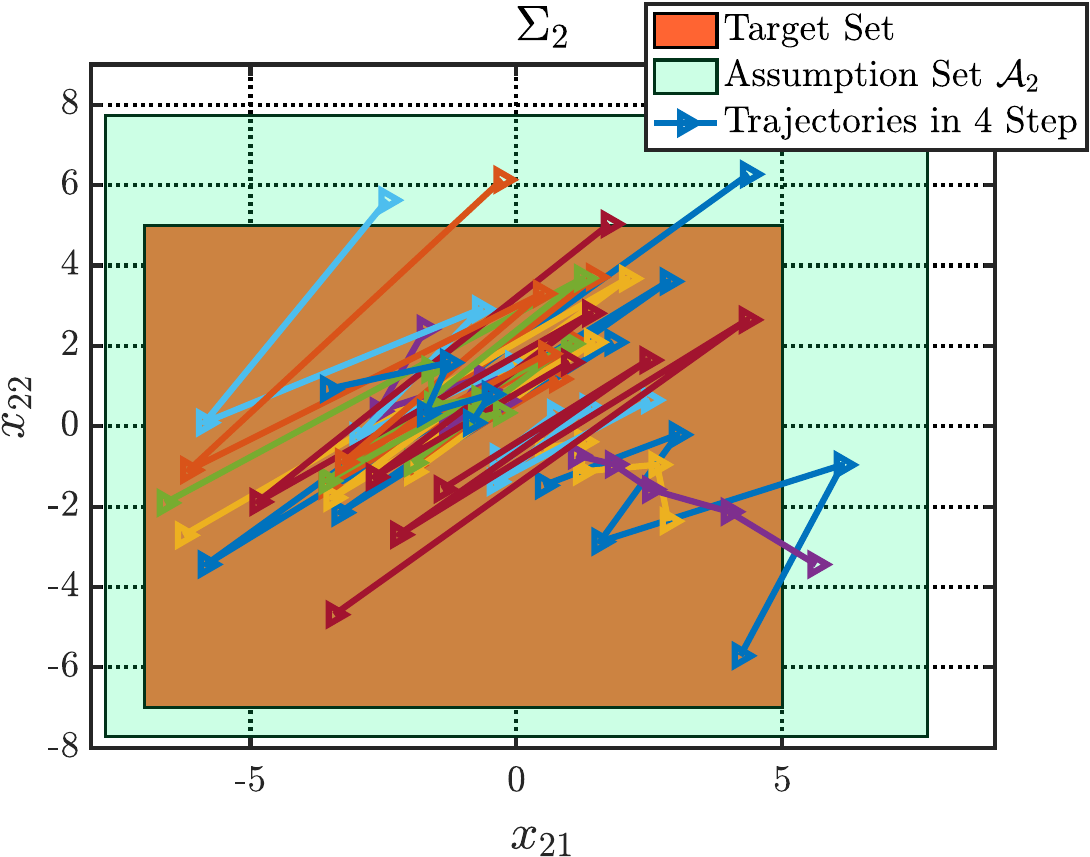}
    \caption{Assumption sets calculated by Algorithm~\ref{Alg: AG }, and 15 trajectories starting from assumption sets and reaching the target set eventually in step~4.}
    \label{fig: Ex FR}
\end{figure}
\end{example}

\begin{remark}
Although the examples presented above utilize identical specification types across all subsystems, the proposed framework is not limited to identical specifications. For instance, one subsystem might be required to satisfy a finite-horizon safety specification, while another may need to meet an exact-time or finite-horizon reachability specification. The algorithms developed in this work can naturally extend to these situations, as the resilience metric is defined in relation to the specific local specification for each subsystem.
\end{remark}
\subsection{Nonlinear Systems} \label{sec com nonl}
To compute AG contracts for nonlinear interconnected subsystems in \eqref{eq: dt subsys}, the resilience metric can be computed using 
\cite[Section \rom{5}.B]{saoud2024temporal}, which is based on Satisfiability Modulo Theories (SMT).
Using SMT solvers, \emph{e.g.}, $\mathsf{dReal}$, we iteratively refine \( g_{\psi}(x) \) by checking the feasibility of system trajectories under increasing disturbance levels \cite{gao2013dreal}.
For AGC synthesis, we substitute the nonlinear resilience computation into Algorithms~\ref{Alg: AG } and {\ref{Alg: AG FS 2}}, where each iteration updates the assumptions \( \mathcal{A}_i \) for interconnected subsystems. The iterative contract computation ensures that the final assumption sets account for interconnection effects, leading to a compositional verification framework for nonlinear systems.
\section{Case Study: DC Microgrid}
\label{sec:case_study}
To show the practicality of our approach, we implement the proposed algorithm on a nonlinear DC Microgrid. We represent a microgrid model adopted from \cite{saoud2021compositional} as a directed graph $\mathbb{G}(\mathcal{N}, \mathcal{E},\beta)$, where $\mathcal{N}$ is the set of $n$ nodes (buses for power units), $\mathcal{E}$ is the set of $t$ edges (transmission lines), $\beta \in \mathbb{R}^{n \times t}$ is the incidence matrix that captures the graph's topology. The Laplacian matrix is defined as $\mathsf{L} := \beta G_T \beta^{\top} \in \mathbb{R}^{n \times n}$, with $G_T := \operatorname{diag}(G_e) \in \mathbb{R}^{t \times t}$, where $G_e$ represents the conductance of each edge. $\mathcal{N}_S$ denotes nodes associated with controllable power units (sources) with $m$ nodes, while $\mathcal{N}_L$ includes nodes related to non-controllable power units (loads), totaling $n - m$ nodes.

The interconnected dynamics of the voltage buses are
\begin{equation}
    C \dot{V}=-(\mathsf{L}+G) V+\sigma, \label{eq dc micro dynamic}
\end{equation}
where $V:=\operatorname{col}\left(v_i\right) \in \mathbb{R}_0^{+n}$ denotes the collection of (positive) bus voltages, $\sigma:=\operatorname{col}\left(\sigma_i\right) \in \mathbb{R}^n$ denotes the collection of input currents and $C:=\operatorname{diag}\left(C_i\right) \in \mathbb{R}^{n \times n}, G:=\operatorname{diag}\left(G_i\right) \in \mathbb{R}^{n \times n}$ are matrices denoting the bus capacitances and conductances. Input currents are given by
\begin{equation}
\sigma_i=\left(\left(1-b_i\right) P_i+b_i u_i\right) / v_i, \quad i \in \mathcal{N},
\label{eq dc micro input}
\end{equation}
with control input $u_i \in \mathcal{U}_i$, where $\mathcal{U}_i:=\left[\underline{u}_i, \bar{u}_i\right] \subset \mathbb{R}^{+} ; b_i \in$ $\{0,1\}$, where $b_i=1$, if $i \in \mathcal{N}_S$ and $b_i=0$ otherwise; and $P_i$ is a bounded time-varying demand $P_i \in \mathcal{P}_i=\left[\underline{P}_i, \overline{P}_i\right]$. By replacing \eqref{eq dc micro dynamic} with \eqref{eq dc micro input}, the overall system can be rewritten in compact form as
\begin{equation}
\dot{V} \in f(V, u)=-C^{-1}\left[(\mathsf{L}+G) V+\left[\begin{array}{c}
u \\
\mathcal{P}
\end{array}\right] \oslash V\right] \label{eq dc micro sys}
\end{equation}
with state vector $V \in \mathbb{R}_0^{+n}$; control input $u \in \mathcal{U}$, where $\mathcal{U}:=\prod_i \mathcal{U}_i$; disturbance input $\mathcal{P}:=\prod_i \mathcal{P}_i$; and where $\oslash$ denotes the element-wise division of matrices. The continuous-time dynamics in \eqref{eq dc micro sys} can be discretized with an appropriate sampling time $\tau$ to 
\begin{equation}
    {V}(k+1) =V(k)- \tau C^{-1}\left[(\mathsf{L}+G) V(k)+\left[\begin{array}{c}
u \\
\mathcal{P}
\end{array}\right] \oslash V(k)\right].
\end{equation}

\begin{figure}
    \centering
    \includegraphics[width=0.6\linewidth]{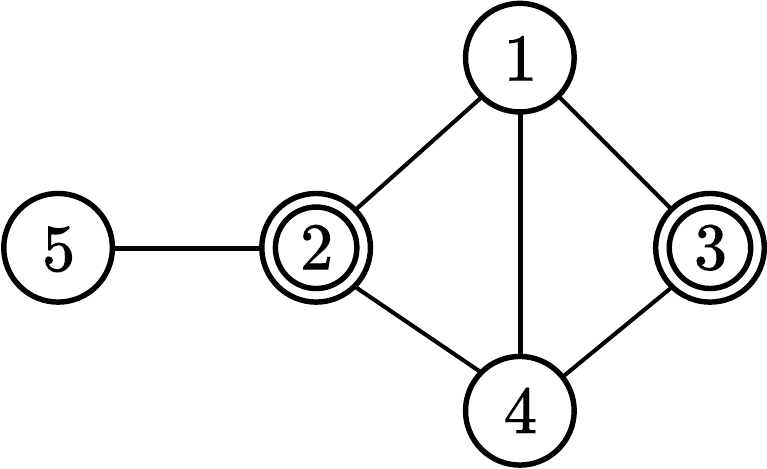}
    \caption{Five units architecture used for the simulations. Circles correspond to loads, and sources are denoted by double circles. Solid lines denote the transmission lines.}
    \label{fig:5unit micro}
\end{figure}

It is required that the voltage $V$ of the system need to be close to the nominal value $V_{\text {nom }}>0$ with deviation of $\delta>0$ at a specific time step $N$, which can be defined as a target set $\left[V^{\text {nom }}-\delta, V^{\text {nom }}+\delta\right]^n$. A five-terminal DC microgrid is depicted in Fig.~\ref{fig:5unit micro}. In this setup, we assume that units 2 and 3 are equipped with a primary control layer, while the remaining three units—units 1, 4, and 5—represent loads with demand that varies steadily around a constant power reference. These loads can therefore be interpreted as constant power loads that are impacted by noise.

The considered bus parameters are $C_1=2.2 \mu \mathrm{~F}, C_2=$ $1.9 \mu \mathrm{~F}, C_3=1.5 \mu \mathrm{~F}, C_4=C_5=1.7 \mu \mathrm{~F}$, and the network parameters are $G_{12}=5.2 \Omega^{-1}, G_{13}=4.6 \Omega^{-1}, G_{14}=4.5 \Omega^{-1}$, $G_{24}=6 \Omega^{-1}, G_{25}=3.1 \Omega^{-1}, G_{34}=5.6 \Omega^{-1}, G_{15}=G_{23}=$ $G_{35}=G_{45}=0 \Omega^{-1}$. The system is supposed to reach the region with grid nominal voltage $V^{\mathrm{nom}}= 450 \mathrm{~V}$ and $\delta= 0.1 V^{\mathrm{nom}}$ at time step $N=15$. Hence, we define the error dynamics as 
\begin{align}
    \Sigma: E(k+1) =&E(k)- \tau C^{-1}\bigr[(\mathsf{L}+G) (E(k)+V^{\mathrm{nom}}) \nonumber\\
    & +\begin{bmatrix}
        u \\
\mathcal{P}
\end{bmatrix} \oslash (E(k)+V^{\mathrm{nom}}))\bigr], \label{error dynamic}
\end{align}
where $E:=\operatorname{col}\left(e_i\right) \in \mathbb{R}_0^{+n}$. In this case, the target set for the error dynamics will be $\Gamma = [-\delta,\delta]^5$, and the requirement of reaching the target set at time step $N=15$ can be translated to an exact-time reachability specification as $\psi = \nex^N\Gamma$. It can be decomposed into the specifications $\psi_i = \nex^{15}\Gamma_i$ for each subsystem, in which $\Gamma_i = [-45,45]$. We consider a scenario in which unit 5 is connected to the grid. We assume that the load power demands for units 1, 4, and 5 are the same, and these units are demanding 0.3 kW each. Source power injections are positive and both are limited to 8 kW. Inputs $u_2$ and $u_3$ serve as feedback controllers, with the gain adjusted to stabilize the system at the nominal voltage. 

\begin{figure}
    \centering
    \includegraphics[width=0.49\linewidth]{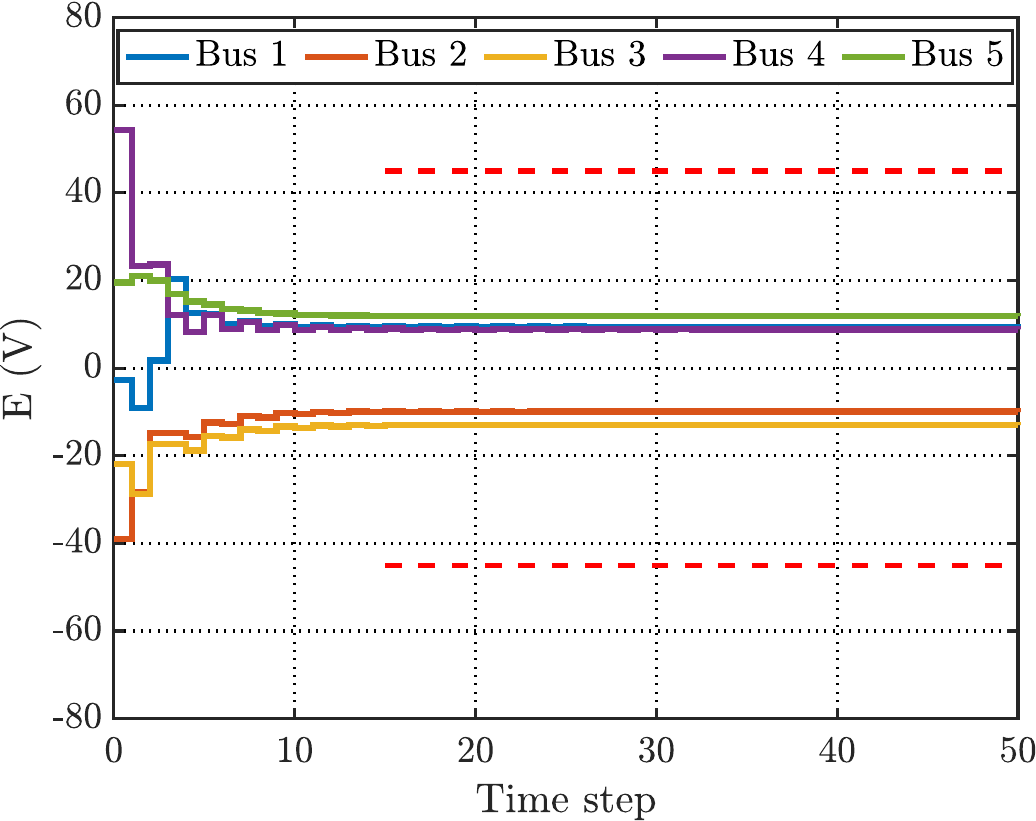}
    \includegraphics[width=0.49\linewidth]{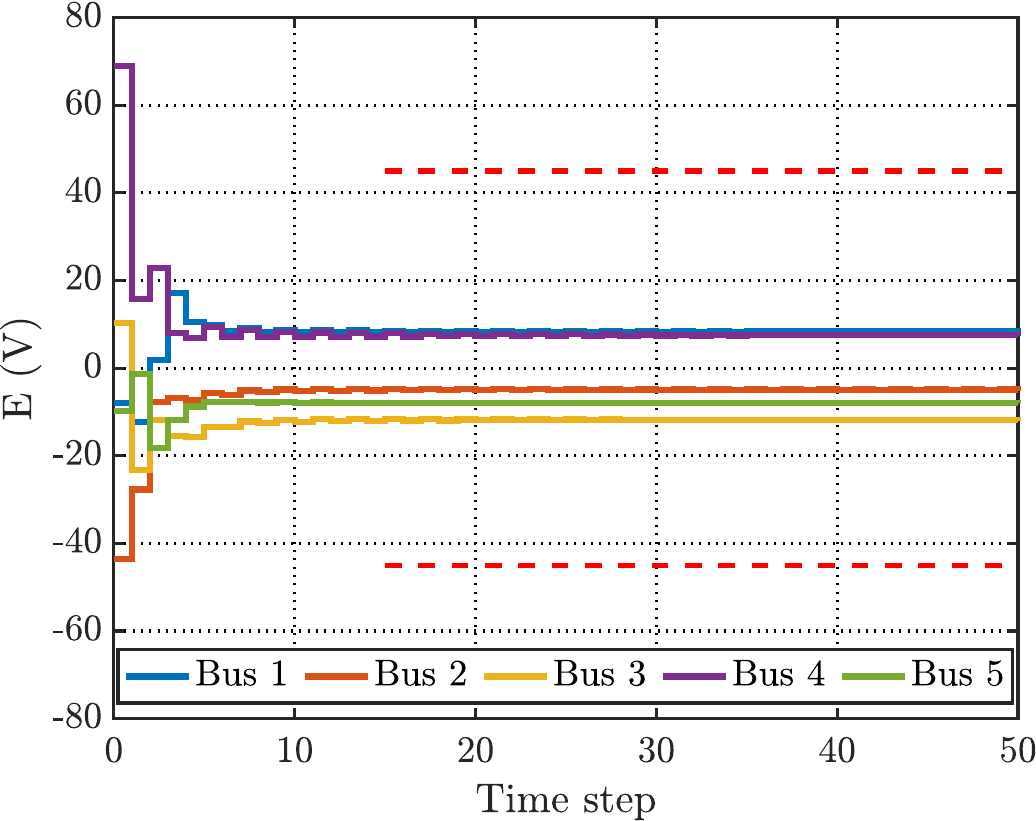}
    \includegraphics[width=0.49\linewidth]{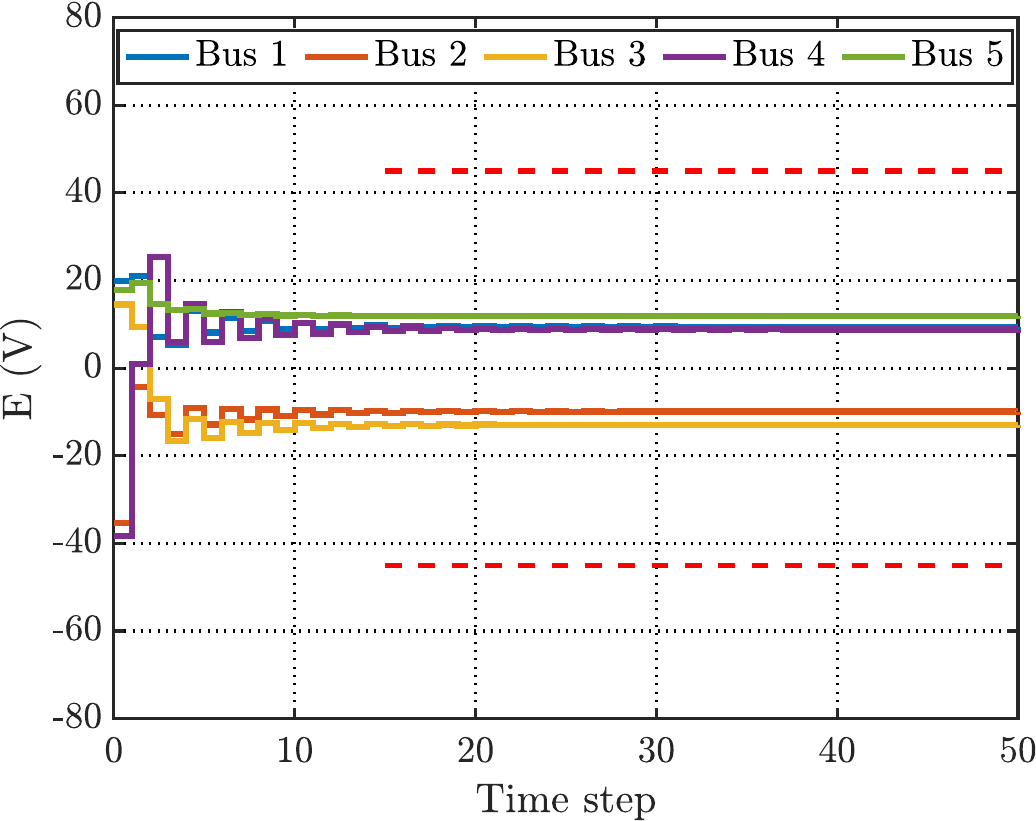}
    \includegraphics[width=0.49\linewidth]{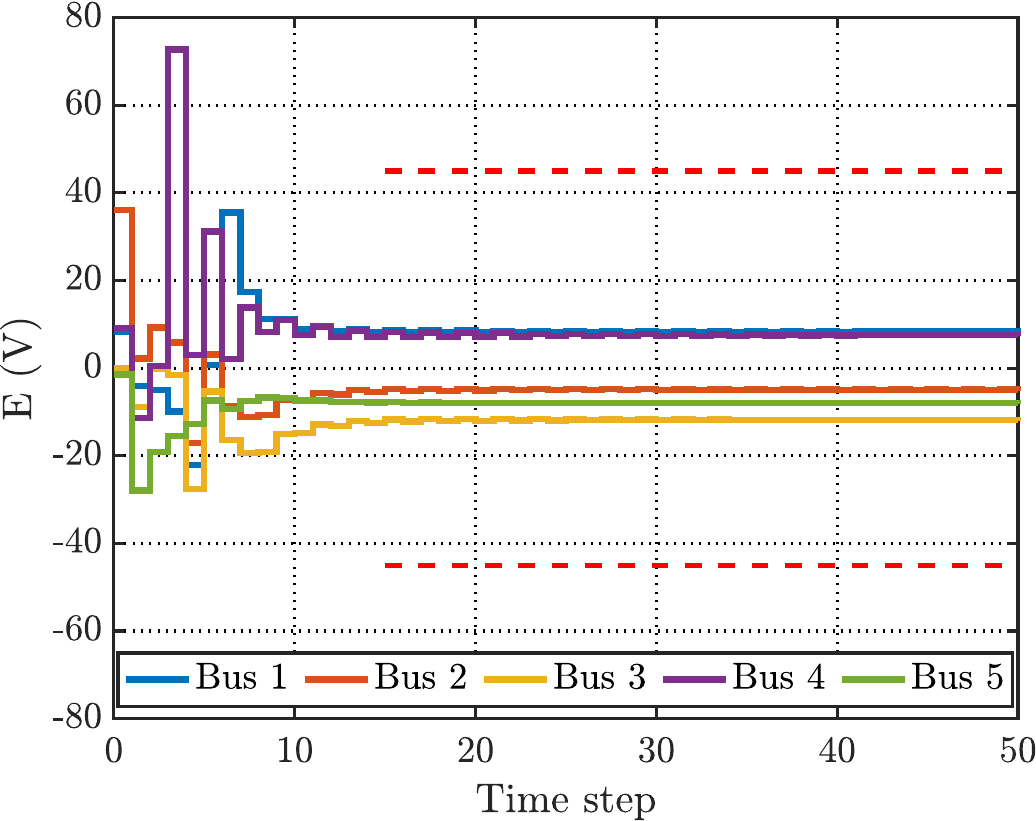}
    \caption{Four random realizations of assumptions on $E$ calculated using Algorithm~\ref{Alg: AG FS 2} to satisfy the specifications $\psi_i = \nex^{15} \Gamma_i$, $\forall i \in \mathbb{N}_5$.}
    \label{fig: dc micro}
\end{figure}

We apply Algorithm~\ref{Alg: AG FS 2} to find the assumptions $\mathcal{A}_i$ that guarantee the satisfaction of specifications $\psi_i = \nex^N \Gamma_i$ for all $i \in \mathbb{N}_5$. We use the SMT-based approach and \(\mathsf{dReal}\) in step 3 of the algorithm as explained in Section~\ref{sec com nonl}. We find the following assumptions for each $\Sigma_i$: $\mathcal{A}_1 = [-36.33,36.33]$, $\mathcal{A}_2 = [-50.24,50.24]$, $\mathcal{A}_3 = [-51.17,51.17]$, $\mathcal{A}_4 = [-75.66,75.66]$, $\mathcal{A}_5 = [-24.73,24.73]$. To demonstrate the correctness of our results, we randomly select four initial values, \( e_i \), for \( i \in \{1, 2, \ldots, 5\} \) and apply them to the interconnected system. As shown in Fig.~\ref{fig: dc micro}, the error in each subsystem converges to the interval \([- 45, + 45]\) by time step 15 and the finite safety specifications defined in the algorithm are satisfied. Thus, the voltages \( V \) will converge to the set \( \Gamma \) at $N=15$. 

\section{CONCLUSIONS AND FUTURE WORKS}
\label{se:conclusions}
In this paper, we proposed a contract-based framework for computing AGCs for interconnected systems using the concept of resilience metric. By integrating resilience metrics into an iterative refinement process, we systematically developed AGCs that ensure compliance with temporal logic specifications for each subsystem. Our framework is applicable to both linear and nonlinear discrete-time systems and supports finite-horizon specifications. The proposed approach maintains maximality for two subsystems. Furthermore, we demonstrate the framework's ability to generalize to multiple subsystems through weighted interconnection reasoning. 
To demonstrate the effectiveness of our approach, we implemented the proposed algorithm on linear case studies and the nonlinear DC Microgrid.
Future work will focus on extending the proposed approach from verification to controller synthesis, aiming to design controllers for each component that guarantee the satisfaction of a given assume-guarantee contract.
%%%%%%%%%%%%%%%%%%%%%%%%%%%%%%%%%%%%%%%%%%%%%%%%%%%%%%%%%%%%%%%%%%%%%%%%%%%%%%%%
% \section{ACKNOWLEDGMENTS}
% 
%%%%%%%%%%%%%%%%%%%%%%%%%%%%%%%%%%%%%%%%%%%%%%%%%%%%%%%%%%%%%%%%%%%%%%%%%%%%%%%%
%%%---------- Ref. ----------%%%
\bibliographystyle{ieeetr}
\bibliography{REF.bib}
\appendix
\subsection{Proofs} \label{sec prf}
\subsubsection{Proof of Theorem \ref{thm converge}}
Suppose Algorithm~1 terminates after $j^\star$ iterations and outputs contracts 
$\mathcal{C}_1=(\mathcal{A}_1,\mathcal{G}_1)$ and $\mathcal{C}_2=(\mathcal{A}_2,\mathcal{G}_2)$, 
with $\mathcal{A}_i := \mathcal{A}_i^{j^\star}$ and $\mathcal{G}_i := \psi_i^{j^\star}$ for $i\in\{1,2\}$.

(1) By construction of the resilience metric, for all disturbance trajectories 
$w_i: \ \in \mathbb{N}_N \rightarrow W_i$, the trajectories of subsystem $\Sigma_i$ satisfy the specification 
$\psi_i^{j^\star}$. Hence, $\Sigma_i \models_c (\mathcal{A}_i^{j^\star},\psi_i^{j^\star})$.  

(2) At termination, the equalities $\psi_i^{j^\star+1}=\psi_i^{j^\star}$, $i\in\{1,2\}$, imply that 
$\psi_1^{j^\star}\subseteq \phi_1^{j^\star} = \{x_1 \in X_1^N | C_{2,1}x_1 \vDash \mathcal{A}_2^{j^*}\}$ and $\psi_2^{j^\star}\subseteq \phi_2^{j^\star}$, 
so that whenever $\Sigma_1 \models \psi_1^{j^\star}$, the induced input trajectory $w_2=C_{2,1}x_1: \mathbb{N}_N \rightarrow W_2$ the trajectory $x_1: \mathbb{N}_N \rightarrow X_1$ of the system $\Sigma_1$, we will have that $w_2 =C_{2,1}x_1 \vDash \mathcal{A}_2^{j^\star}$. In addition, whenever $\Sigma_2 \models \psi_2^{j^\star}$ the induced input trajectory $w_1=C_{1,2}x_2: \mathbb{N}_N \rightarrow W_1$ the trajectory $x_2: \mathbb{N}_N \rightarrow X_2$ of the system $\Sigma_2$, we will have that $w_1 =C_{1,2}x_2 \vDash \mathcal{A}_1^{j^\star}$.

Combining (1) and (2), it follows that both subsystems satisfy their contracts with consistent 
assumptions and guarantees, i.e. $\Sigma_i \models_c \mathcal{C}_i$ for $i\in\{1,2\}$. 
Therefore, the pair $(\mathcal{C}_1,\mathcal{C}_2)$ forms a solution to Problem~\ref{prblm} for 
two interconnected subsystems. \hfill$\blacksquare$

\smallskip
\subsubsection{Proof of Theorem \ref{thm monotonic}}
Fix an iteration index $j \in \mathbb{N}$. In Algorithm~\ref{Alg: AG }, a resilience metric $\varepsilon_1^j$ is computed for subsystem $\Sigma_1$ and defines the assumption on its internal input as $\mathcal{A}_1^j = \square^N \Omega_{\varepsilon_1^j}(0) \subseteq W_1^N$. Using the interconnection relation, this is translated into a finite-horizon safety specification on $x_2$, denoted by $\phi_2^j$. Then, the update of $\Sigma_2$'s guarantee is
\[
\psi_2^{j+1} \;=\; \psi_2^{j} \wedge \phi_2^{j} \;\subseteq\; \psi_2^{j}.
\]
Next, a resilience metric $\varepsilon_2^j$ is calculated for subsystem $\Sigma_2$, leading to an assumption $\mathcal{A}_2^j = \square^N \Omega_{\varepsilon_2^j}(0) \subseteq W_2^N$, which is translated into a safety specification $\phi_1^j$ on $x_1$. The update for $\Sigma_1$ is
\[
\psi_1^{j+1} \;=\; \psi_1^{j} \wedge \phi_1^{j} \;\subseteq\; \psi_1^{j}.
\]
Therefore, at every iteration the sequences $\{\psi_1^j\}_{j\in\mathbb{N}}$ and $\{\psi_2^j\}_{j\in\mathbb{N}}$ are monotonically non-increasing with respect to set inclusion. Equality holds if the new constraint $\phi_i^j$ already contains $\psi_i^j$; otherwise, the refinement is strict. \hfill$\blacksquare$

\smallskip
\subsubsection{Proof of Theorem \ref{thm convergence l subsys}}
Suppose Algorithm~\ref{Alg: AG FS 2} terminates at iteration $j^\star$ and for each $i\in\mathbb{N}_L$, the contract set is $\mathcal{C}_i=(\mathcal{A}_i,\mathcal{G}_i)$ with $\mathcal{A}_i=\square^{N}\ball_{\varepsilon_i^{\,j^\star}}(0)$ and $\mathcal{G}_i=\psi_i^{\,j^\star}$. By definition of the resilience metric, any input sequence $w_i\in\mathcal{A}_i$ guarantees that $\Sigma_i$ satisfies $\mathcal{G}_i$, i.e., $\Sigma_i\models_{\mathfrak c}\mathcal{C}_i$. At each iteration, the radius $\varepsilon_i^{\,j}$ is split across neighbors $\ell\in\mathcal{L}_i$ with weights $\lambda_\ell^i\ge 0$, $\sum_{\ell\in\mathcal{L}_i}\lambda_\ell^i=1$, producing neighbor constraints that are conjoined to $\psi_\ell^j$. Termination means $\psi_\ell^{\,j^\star+1}=\psi_\ell^{\,j^\star}$ for all $\ell$, so each neighbor’s guarantee already implies all received constraints; hence if $\Sigma_\ell\models \psi_\ell^{\,j^\star}$ then $C_{i,\ell}x_\ell\in \ball_{\lambda_\ell^i \varepsilon_i^{\,j^\star}}(0)$. For the interconnection $w_i=\sum_{\ell\in\mathcal{L}_i}C_{i,\ell}x_\ell$, the triangle inequality and $\sum_\ell \lambda_\ell^i=1$ yield
$\lVert w_i(k)\rVert_\infty \le \sum_{\ell}\lambda_\ell^i \varepsilon_i^{\,j^\star}=\varepsilon_i^{\,j^\star}$ for all $k$, which implies that $w_i\in \square^{N}\ball_{\varepsilon_i^{\,j^\star}}(0)=\mathcal{A}_i$. Thus, neighbor guarantees make all assumptions hold, and by local correctness, each $\Sigma_i$ satisfies $\mathcal{G}_i$ and solves Problem~\ref{prblm}. \hfill$\blacksquare$
\end{document}